\documentclass[superscriptaddress,preprint, showpacs]{revtex4-1}

\usepackage{amsmath}
\usepackage{amssymb}
\usepackage{bbold}
\usepackage{mathrsfs}
\usepackage{graphicx, psfrag}
\usepackage[centerlast]{caption}
\usepackage{float}
\usepackage{subfig}
\usepackage{tikz}
\usepackage{pgfplots}
\usepackage[colorlinks=true, citecolor=blue, urlcolor = blue, linkcolor= blue, bookmarks=true]{hyperref}
\usepackage{epstopdf}
\usepackage{multirow}
\usepackage{tabularx} 
\usepackage[english]{babel}
\usepackage[utf8]{inputenc}
\usepackage{fancyhdr}

\def \bb{\bibitem}
\def \nn{\nonumber}
\def \bs{\boldsymbol}
\def \( {\Big (}
\def \) {\Big )}

\def \bef{\begin{figure}[H]}
\def \eef{\end{figure}}
\def \beq{\begin{equation}}
\def \eeq{\end{equation}}
\def \bse{\begin{subequations}}
\def \ese{\end{subequations}}
\def \bea{\begin{eqnarray}}
\def \eea{\end{eqnarray}}

\begin{document}

\title{Vortex Dynamics of Rotating Bose-Einstein Condensate of Microcavity Polaritons}

\author{Bikash Padhi}
\email{bpadhi2@illinois.edu}
\affiliation{Department of Physics, Indian Institute of Technology Delhi, New Delhi 110016, India}
\author{Romain Duboscq}
\email{romain.duboscq@math.univ-toulouse.fr}
\affiliation{Institut de Math\'ematiques de Toulouse, UMR 5219, Universit\'e Paul Sabatier Toulouse 3, 118 Route de Narbonne, 31062 Toulouse Cedex 9, France}
\author{Ankita Niranjan}
\affiliation{Department of Physics, Indian Institute of Technology Delhi, New Delhi 110016, India}
\author{Ravi K. Soni}
\affiliation{Department of Physics, Indian Institute of Technology Delhi, New Delhi 110016, India}

\begin{abstract}
In this work we perform a numerical study of a rotating, harmonically trapped, Bose-Einstein condensate of microcavity polaritons. An efficient numerical method (toolbox) to solve the complex Gross-Pitaevskii equation is developed. Using this method, we investigate how the behavior of the number of vortices formed inside the condensate changes as the various system parameters are varied. In contrast to the atomic condensates, we show, there exists an (experimentally realizable) range of parameter values in which all the vortices can be made to vanish even when there is a high rotation. We further explore how this region can be tuned through other free parameters and also discuss how this study can help to realize the synthetic magnetic field for polaritons and hence paving the way for the realization of the quantum Hall physics and many other exotic phenomena.
\end{abstract}

\pacs{42.65.Sf, 71.36.+c, 42.55.Sa, 47.32.-y}

\maketitle

\section{Introduction}

Ever since the observation of the Bose-Einstein condensate (BEC) of ultracold
($\sim$100 nK) \textsuperscript{87}Rb atoms at NIST-JILA, in 1995 \cite{BEC}, the field of cold atoms has grown enormously in the last two decades \cite{MaciejBook}. The primary factor that still remains a great challenge for experimentalists in cold atom physics is the extremely low temperature required to perform experiments. Hence soon after this breakthrough many other counterparts of the Rubidium BEC were sought for, in alternative bosonic systems. In 2006 Kasprzak {\it et al.} reported the first observation of a BEC of microcavity polariton (MCP) \cite{Kasprazak}, which was theoretically proposed in \cite{Poras}. The microcavity polariton condensate (MPC) was realized only at a few Kelvins. MCP are bosonic quasi-particles formed as a superposition of semiconductor excitons and photons trapped inside a microcavity \cite{DengRMP}. It can be shown that  \cite{Perrin}, when the exciton and photon are in a strong coupling regime, then the internal structure of the polariton can be ignored and an effective one-particle description can be used. 

 The reason for having such a high critical temperature for the MPC is, when the in-plane wave vector vanishes ($\bs k = 0$), the effective mass of the MCP is about three orders of magnitude smaller than that of the bare quantum well exciton. Recent experiments on polariton BEC have achieved even higher temperatures \cite{Lagoudakis, New}. This increase in critical temperature in MPCs paved the way for a lot of theoretical and experimental works. In particular, because of their finite life-time, which can be experimentally controlled \cite{Bloch}, the MPC are the best candidates for studying dynamic condensates \cite{DengTEQM, PumpI, Vortices, cGPE, KeelingsRot}. These studies have also resulted in various technological advancements \cite{Bloch, New, Perrin}. A review of recent progress in this area can be found in \cite{DengRMP}. 

In this work, we confine ourselves to the study of the steady state situation, i.e., we ignore all the time evolutions and assume the system has achieved a (dynamic) equilibrium state. This dynamical nature of a polariton BEC is an artifact of the competition between various energy scales, namely, inter-particle interaction, kinetic energy, cavity pumping, particle decay. The effects of inter-particle interaction has been studied in \cite{Interaction}. The effects of pumping and decay in these condensates have also been the subject of several recent works \cite{PumpI}. However, these two (tunable) parameters play a very crucial role in deciding the steady state of the condensate. In our work we focus on how the vortex formation in a MPC \cite{NucleationL} is affected by these parameters. Particularly, we search for a region of parameters where the vortices can be made to vanish, even when the condensate is rotating with a very high angular speed. The physical implication of such a parameter regime would be to facilitate the study of synthetic magnetic field on MCP systems \cite{MCPGauge, Dalibard}. We postpone the introduction to synthetic magnetic field until Sec. \ref{sec:Theory}.

\begin{figure}
\centering
\includegraphics[scale=0.5]{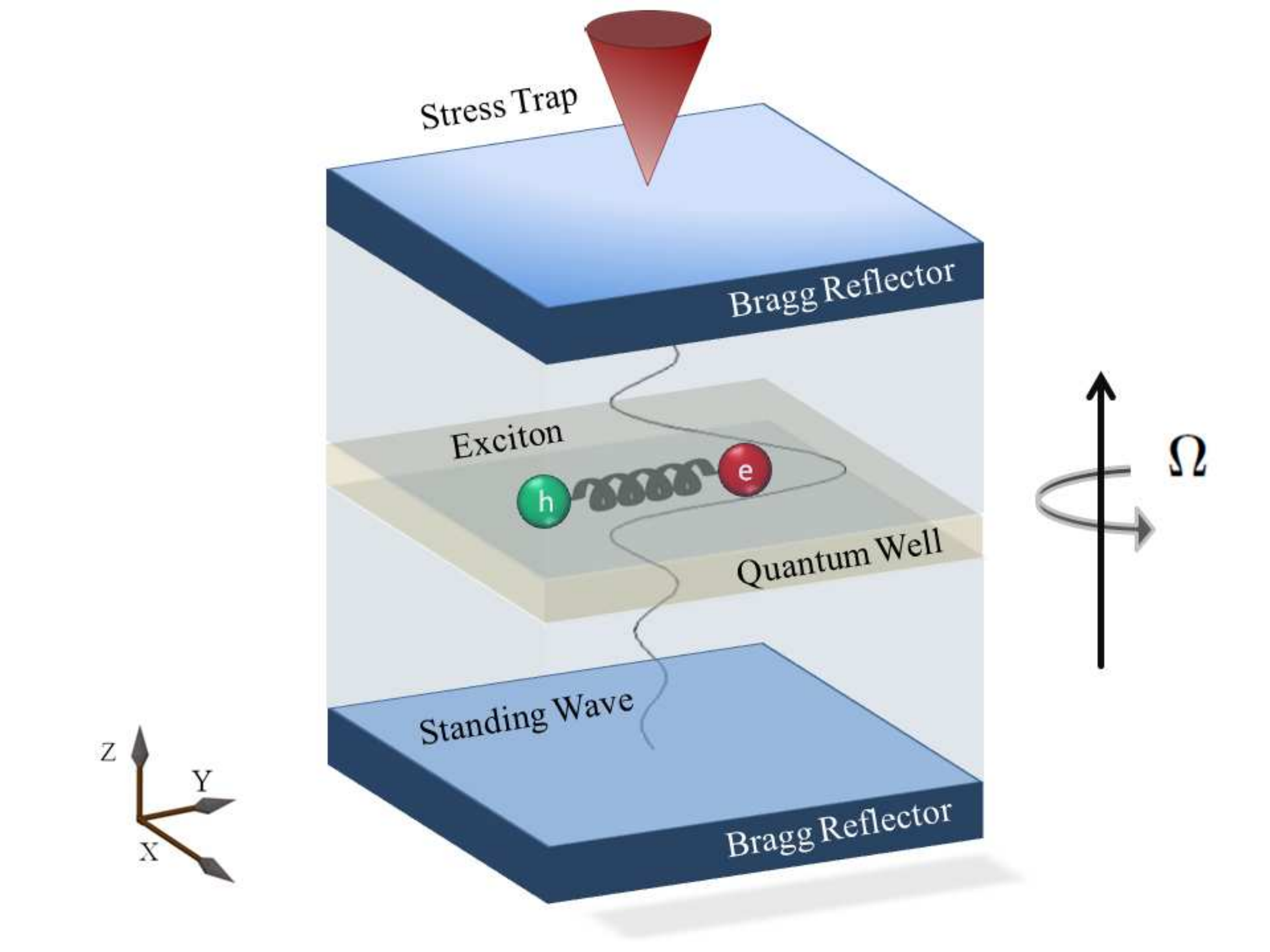}
\caption{Schematic of the model: The semiconductor microcavity which consists of two distributed Bragg reflectors. The semiconductor material used inside (in the middle of the cavity) has a quantum well which traps the excitons, i.e., a bound electron (e) and hole (h) pair. The semiconductor excitons get strongly coupled to the cavity photon to form polaritons. A stress trap is used to generate a harmonic potential \cite{Trap}. The trap is created by the tip of a pin (red inverted cone). Finally, the entire system is made to rotate by an angular speed of $\Omega$ about the $z$-axis.}
\label{fig:Schematic}
\end{figure}

{\it The Model}: We consider a MPC in the superfluid phase. The condensate is trapped inside a harmonic potential trap. The trap can be realized in many ways as described in \cite{Trap, ChenHarmonic}. Here we have chosen a method described in \cite{Trap}, which uses the tip of a pin to realize a harmonic trap. The polaritons are incoherently injected into the cavity (along the $z$-axis)  and they form an inhomogeneous condensate inside it. The size of such a condensate is decided by the spot size of the pumping or illuminating laser. However, in this work, we do not address the issue of how the spot size of the pump laser will affect the MPC. We mechanically rotate the entire system in a specific frequency, $\Omega$. As an alternative way of achieving rotation, a Laguerre-Gauss mode of laser beam with a non-zero orbital angular momentum can also be used \cite{LGBeam}. The stirring time is kept shorter than typical vortex nucleation time, also the choice of the pump power is kept low enough to avoid  spontaneous vortex formation \cite{ChenVortices, Vortices}. In most cases, polaritons have a lifetime shorter than the cooling time and the system remains in quasi-equilibrium. However, Deng {\it et al.} \cite{DengTEQM} showed that above a certain threshold pump power and a large positive detuning, the thermalization time can be shorter than the polariton lifetime and for a time interval the system can be brought to thermal equilibrium with the lattice. We restrict ourselves to such a regime of pump power.

We also assume the system is under a critical rotation, i.e. when the trap frequency and the rotation frequency are very close to each other. We will see this is a condition which has to be satisfied in order to engineer artificial magnetic field. The paper is organized in the following way: in the Sec. \ref{sec:Theory} we provide the theoretical background of synthetic magnetic field, the theory of MCP and discuss the complex Gross-Pitaevskii equation (cGPE). Then in the Sec. \ref{sec:Numerical}, we describe the numerical procedures used to find out the steady state solutions of the cGPE. In Sec. \ref{sec:ResultsST} we discuss how these steady state of the system are affected by various control parameter. In Sec. \ref{sec:ResultEvo}, we discuss how the time-evolution (of these steady states) depends on the parameters, and then we conclude.

\section{Synthetic gauge field for MCP}
\label{sec:Theory}

As compared to the conventional solid state systems, the high degree of control in case of cold atomic systems or MCP systems allows us to verify and study a broad range of physics, including many exotic quantum phenomena. A phenomenon with perennial interests is motion of a charged particle in a magnetic field. However, both cold atomic systems and the MCP systems are neutral systems and the electromagnetic (gauge) field can only be coupled to a charge particle. Nevertheless, now it is well established that similar physics (effective theories) can also be simulated even with neutral particles \cite{Zee}. Various methods to generate synthetic or artificial gauge field are discussed in \cite{Dalibard}. Using these methods for cold atomic system physicists have successfully simulated many exotic quantum phenomena, such as Hall effect \cite{Hall}, Landau levels \cite{Padhi}, Hofstadter butterfly \cite{HBloch} structure and many more. Here we briefly describe one of the simplest method to engineer synthetic magnetic field for MCP, i.e. by rotating the system \cite{NucleationM}. 

The single particle Hamiltonian of a harmonically trapped quantum particle under rotation can be written as
\bea
\hat{H} &=& \frac{|{\bf {p}}|^2}{2m} + \frac{1}{2}m\omega_{\perp}^2({x}^2 + {y}^2) - {\bf \Omega .{r}\times {p}} \nn \\
&=& \frac{1}{2m}\Big({\bf|p|^2} - 2m{\bf p.\Omega\times r}  \Big ) + \frac{1}{2}m\omega_{\perp}^2({x}^2 + {y}^2) .
\eea
So clearly when the rotation frequency ($\Omega$) is very close to the trap frequency ($\omega_\perp$) the resulting Hamiltonian becomes equivalent to a charged particle in a magnetic field. This analogy provides a quasi charge of $q = m$ and a synthetic magnetic field of strength $\bs B = 2\Omega \hat{z}$. One must be cautioned here that the rotation does not cause emergence of any real magnetic field but the effective Hamiltonian of the system just becomes equivalent to a charge moving in a magnetic field. Although the polaritons are neutral particles, but we can mimic the dynamics of a charged particle by realizing such artificial field. This idea can also be well extended to many-particles {\it provided} there exists no singularity (or, vortex) in the many-particle wave function. This is the key motivation of our work: to find out an experimentally viable range of parameter regime, where we can avoid singularities or vortices.

The dynamics of the quantum fluid of polaritons can be studied under mean-field approximation. The Gross-Pitaeveskii equation (GPE) is a mean-field model to deal with the many-body problems \cite{Stringari}. For temperature below certain critical temperature, the classical field or the macroscopic wave function of the condensate, $\Psi(\bs r, t)$ is used to replace the field operator. Similarly, the dynamics of a weakly (repulsively) interacting polariton superfluid can be described by the complex GPE (cGPE) \cite{cGPE}. This is a generic mean-field model for non-equilibrium (or, quasi-equilibrium for a given pumping rate) MPCs, considering effects from pumping, dissipation, potential trap, relaxation and interactions \cite{PumpI},
\beq
\imath \hbar \frac{\partial \Psi}{\partial t} = \Big [ -\frac{\hbar^2 \nabla ^2}{2m} + V(\bs r) + g|\Psi|^2 + \imath \hbar \nabla . (\bs \Omega \times \bs r) + \imath (\gamma_{eff} - \Gamma |\Psi|^2) \Big] \Psi. 
\label{cGPE}
\eeq

Since we consider incoherent pumping, $\gamma_{eff} = \gamma - \kappa$ is introduced to describe the net gain; $\gamma$ describes the rate of stimulated scattering of polaritons into the condensate and $\kappa$ describes polariton decay out of the cavity. The  constant $\Gamma$ is introduced to ensure the condensate density saturates and an equilibrium density is attained. $g = 4\pi \hbar^2 a_s/m$ is the effective interaction strength, where $a_s$ is the $s$-wave scattering length (positive for repulsive interaction and negative for attractive
interaction). $V(\bs r) = m(\omega_x^2x^2+ \omega_y^2y^2)/2$ is the external harmonic trap. $m$ is effective-mass of the MCP. Unless stated, we assume $\omega_x = \omega_y = \omega_\perp$. $\Omega$ is the angular frequency of rotation.
In order to derive the time-dependent cGPE in \eqref{cGPE}, one needs to impose the stationary condition 
\bea
\delta && \Big [ - i \hbar \int \bar{\Psi} \frac{\partial	}{\partial t} \Psi d \bs r dt + \int \mathcal{E} dt \Big ] = 0 \quad , \quad \mathcal{E}(\Psi, \overline{\Psi}) = \int \mathcal{L} \,  dr, \nn \\
\mathcal{L} =&& \frac{\hbar^2}{2m}|\nabla \Psi|^2 + V(r)|\Psi|^2 + \frac{g}{2}|\Psi|^4 - \Omega \overline{\Psi}L_z\Psi +  \imath \Big ( \gamma_{eff} - \frac{\Gamma}{2}|\Psi|^2 \Big )|\Psi|^2   \, .
\label{eq:Efunc}
\eea

For studying the steady state of the condensate \cite{Stringari} we consider the time-independent cGPE. For the computation of this classical ground state or "equilibrium state", one needs to minimize the energy functional, $\mathcal{E}(\Psi, \overline{\Psi})$. It must be noted that we work under the normalization constraint, $\|\Psi\|_0^2 : =\int_{\mathbb{R}^d} \bar{\Psi} \Psi d\boldsymbol{x} = 1,$ ($\bar{\Psi}$ is the conjugate field of $\Psi$). This might not be a good approximation for a strongly non-linear regime since the effective mass induced by the non-linearity might cause huge change in the norm  in every computation step, however for a typical experimental scenario (see e.g. \cite{Vortices}) this might be a suitable approximation. 

Keeping all these situations in mind, we proceed to discuss the numerical investigation of the cGPE. Our numerical tool \cite{Antoine2014} is primarily based on \cite{Bao} and references there in. In this toolbox we can tune various free parameters in the problem and minimize the energy functional. In the next section we briefly discuss how the toolbox works. It must be noted that, in a strict sense, this is not equivalent to minimization of free-energy to find the ground state of the system. In fact, since we are considering an open and driven system this minimization may not be defined, since the entropy cost of putting energy in the system is not fixed. So, in our context, what we mean by "minimization" is, given a ground state, we determine how the system evolves to a equilibrium state by computing the energy functional on every iteration and making the functional drop below a certain threshold value. Note, henceforth we refer to the final state as equilibrium state, instead of steady state. This will be addressed in more detail in the following section.


\section{Computation of stationary states}
\label{sec:Numerical}

In this section we describe the procedure to obtain the numerical solution of Eq. \eqref{cGPE}, but before that, we introduce \cite{Kasprazak, Trap, Bao} the scales used for removing dimensions from the problem in Table \ref{tbl:dimension}.
\begin{table}[h!]
\centering
\caption{\label{tbl:dimension} Notations and scaling parameters used. Here $\omega_\perp = \text{min}(\omega_x , \omega_y)$ and $a_0$ is the oscillator length $\sqrt{\hbar/ m \omega_\perp}$. The typical ranges of the parameters are also provided based on the experiments of \cite{Kasprazak, Vortices, Trap}.}
	\begin{tabularx}{\textwidth}{cccc}
		\hline \hline
{\bf Variable/ Parameter}  & {\bf Notation} & {\bf Scale} & {\bf Range}  \\ \hline 
Time & $t$ & $1/\omega_\perp$ & -- \\
Energy & $E$ & $\hbar \omega_\perp$ & -- \\
Length & $x$  & $a_0$ & 1 - 10 \\
Condensate Density & $|\Psi(\bs x, t)|^2$ & $1/a_0^3$ & 0 - 1 \\
Rotation Frequency & $\Omega$ & $\omega_\perp$ & 0 - 1 \\
Interaction strength & $g$ & $a_0^3 \hbar \omega_\perp$ & 0 - 50 \\
Complex interaction & $\Gamma$ & $a_0^3 \hbar \omega_\perp$ & 0 - 50 \\
Effective gain & $\gamma_{eff}$ & $\hbar \omega_\perp$ & 0 - 20\\
		\hline \hline
	\end{tabularx}
\label{Tab:Scale}
\end{table}

After scaling all the variables by appropriate scales we use an initial data which has a gaussian form (since the ground state of an harmonic oscillator is a gaussian wave function). The form of the gaussian is \cite{Bao}
\bea
\Psi^{init} (\bs x) &=& \frac{(1- \Omega)\phi_0(\bs x) + \Omega \phi_1(\bs x)}{||(1- \Omega)\phi_0(\bs x) + \Omega \phi_1(\bs x)||} , \nn \\
\phi_0 (\bs x) &=& \frac{(\omega_x \omega_y)^{1/4}}{\sqrt{\pi}} \exp^{-(\omega_x x^2 + \omega_y y^2)/2} \nn \\
\phi_1 (\bs x) &=& \frac{(\omega_x x + i \omega_y y)}{\sqrt{\pi}} \exp^{-(\omega_x x^2 + \omega_y y^2)/2}.
\eea

For large rotation frequencies ($\gtrsim 0.5$) the initial choice automatically picks up a vortex at the center of the condensate. The end result of the minimization procedure does not depend on the type of initial choice we have, rather such a condition on the initial choice ensures faster convergence towards the final state \cite{Bao}.

{\it The imaginary time method}: There is a great variety of methods to compute stationary states for Gross-Pitaevskii equations. A well-known method is the \textit{imaginary time} method \cite{Adhikari1,ImaginaryBaye,Cerimele1,Chiofalo1,EDWARDSBURNETT,Gammal1}. Mathematically, it consists in a projected gradient applied to the energy associated to the Gross-Pitaevskii equation \cite{BaoDuSISC}. This method is divided in two consecutive steps: a gradient flow and a projection on the constraint space. That is, we iteratively: solve the so-called \textit{imaginary time} Gross-Pitaevskii equation on a time step and then normalize the solution. Considering a uniform time discretization $t_0 = 0 < t_1 < t_2< ...$ with $\delta t = t_{n+1} - t_n$, $\forall n\in\mathbb{N}$, and applying the \textit{imaginary time} method to \eqref{eq:Efunc}, we obtain
\begin{equation}\label{GFDN}
\left\{\begin{array}{l|}
\displaystyle -\hbar \partial_{t} \phi =  \left(\frac{1}{2}\nabla^2 + \Omega L_z - V(\boldsymbol{r}) - g|\phi|^2 - i\left[\gamma_{\textrm{eff}} - \Gamma|\phi|^2 \right]\right)\phi, \ 
 t_{n}<t<t_{n+1},
  \\[4mm]\displaystyle \phi(\boldsymbol{r},t_{n+1})= \lim_{\tau\rightarrow t_{n+1}}\frac{ \phi(\boldsymbol{r},\tau)}{||\phi(\cdot,\tau)||_{0}},
\end{array}\right.
\end{equation}
with an initial data $\phi(\cdot,0) = \phi^0$. It has been shown that, under certain hypothesis, this type of method computes an energy diminishing sequence $(\phi(\cdot,t_n))_{n\in\mathbb{N}}$ \cite{BaoDuSISC}. Therefore, we obtain the following approximation of a stationary state $\phi_{g}$,  $ \phi(\cdot,t_n)\underset{n\rightarrow \infty}{\approx} \phi_{g}.$

However, in practice, we have to fix a stoping criterion. Since the objective is to compute a critical point of the energy \eqref{eq:Efunc}, we set the following criterion based on the evolution of the energy between two iterations, $ |\mathcal{E}(\phi^{n+1},\bar{\phi^{n+1}}) - \mathcal{E}(\phi^n,\bar{\phi^n})| < \varepsilon\delta t,$ where we denote $\phi^n = \phi(\cdot,t_n)$ and with $\varepsilon\ll 1$.
Finally, we have to choose a time and space discretization for the Eq. \eqref{GFDN} in order to efficiently compute this sequence. In \cite{antoine2014robust,bao2006efficient}, the authors show that a Backward Euler time discretization and a pseudo-spectral approximation leads to an efficient scheme called BESP. Moreover, by using Krylov subspaces solvers, the scheme is made fast and robust \cite{antoine2014robust}. 

Based on the above described methods a Matlab toolbox, called GPELab \cite{Antoine2014} is developed. It computes stationary states as well as dynamical solutions of a large class of Gross-Pitaevskii equations and is developed around spectral schemes and includes the BESP scheme \cite{Antoine2014}. Since this code is easy to use and flexible, we choose to compute the stationary states of the cGPE with the help of this toolbox. In Appendix \ref{app:gpelab} we describe it's working in bit more detail. In the following two sections we present our results. First we discuss how the equilibrium state can be controlled by the free parameters and then in the next section, we discuss how the time evolution to these states can be affected by these parameters.

\section{The Equilibrium State}
\label{sec:ResultsST}

The peculiarity of a rotating condensate is presence of vortices inside the condensate, which are the singularities present in the condensate density. This phenomenon is similar to the rotation of \textsuperscript{4}He superfluid. The control over the vortices in a rotating condensate can facilitate investigation of many quantum behaviors on a macroscopic level. As mentioned before, our primary motivation for this work is to understand how the vortex number can be controlled using tunable-parameters, in the equilibrium state ($\partial \Psi / \partial t = 0$). A key observation would be to note that the damping term $i \gamma_{eff}$ can always be removed from the cGPE by the following gauge transformation :
\beq
\Psi(\bs x, t) \rightarrow \Psi(\bs x, t) \exp (i\gamma_{eff} t/ \hbar).
\eeq
Hence the equilibrium state is not going to be affected by $\gamma_{eff}$, however, the evolution to that particular state is going to depend on the value of $\gamma_{eff}$. 
Physically, $\gamma_{eff}$ models the effects of the finite life-time of the polaritons inside the cavity, therefore, it is quite natural that $\gamma_{eff}$ does not bring any difference in the configuration of the equilibrium state, except deciding the time scale of evolution. Hence in this section we only consider the effects of $g, \Omega, \Gamma$. 

\begin{figure}[]
\centering
\subfloat[]{
\includegraphics[width=0.5\textwidth, height=0.4\textwidth]{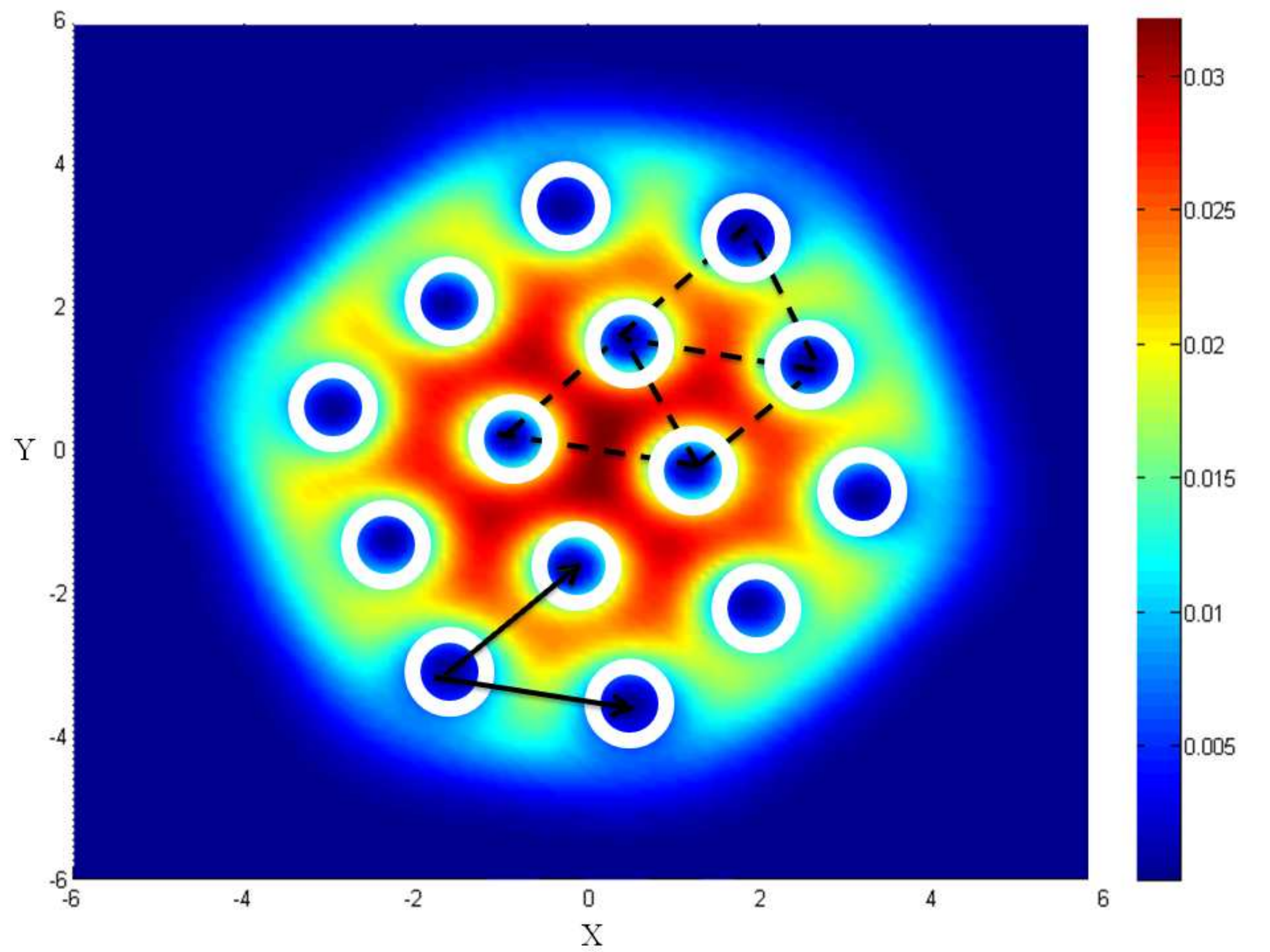} \label{VAmp} }
\subfloat[]{
\includegraphics[width=0.5\textwidth, height=0.4\textwidth]{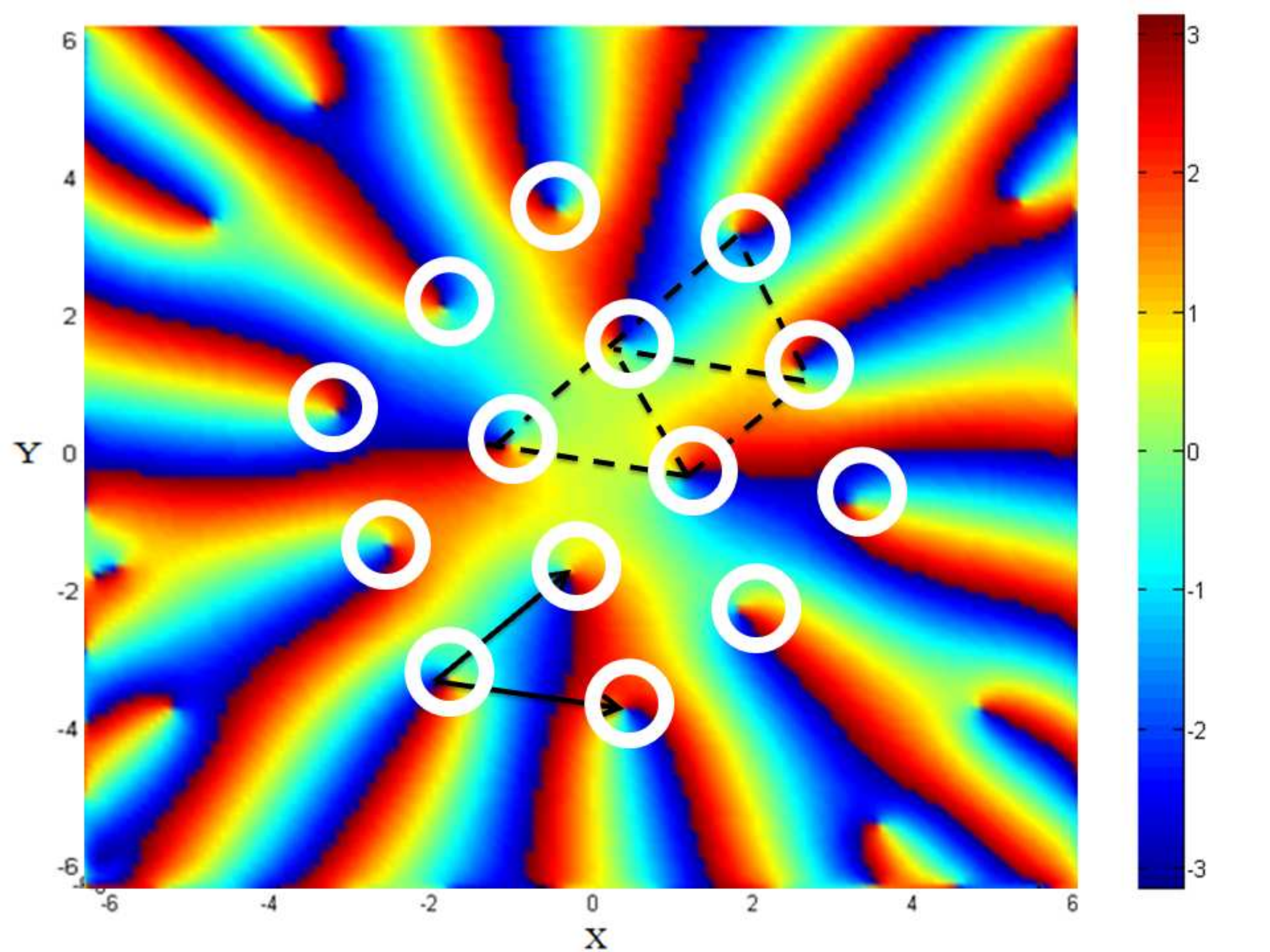} \label{VPhase} }
\caption{The vortices ($N_v = 14$) are found to arrange themselves in a triangular lattice. The two black solid arrows are the lattice vectors. The vortices are marked by white colored rings. Figure (a) is the amplitude of the wave function and (b) is the phase part of the wave function. The color map bar in (a) is a density map, red is highly dense and blue is minimum density. The color map bar in Figure (b) shows variation of angle from $-\pi$ (blue) to $\pi$ (red). This figures corresponds to ($g, \Gamma, \gamma_{eff}, \Omega$) = (50, 0, 20, 0.95).}
\label{Vortices}
\end{figure}

For illustrative purposes, in Fig. \ref{VAmp}, for $(g, \Gamma, \gamma_{eff}, \Omega$) = (50, 0, 20, 0.95), we show the condensate density distribution (with the help of a color map) over the real space. The vortices correspond to the points inside the condensate where the density abruptly vanishes (blue dots with white rings around). The phase of the wave function is depicted in Fig. \ref{VPhase}. We can see the vortices are also singularities in the phase of the wave function since the phase varies from $-\pi$ to $\pi$ (or the other way around) in there vicinity. In the equilibrium state the vortices are always found to arrange themselves in a triangular lattice (the so-called Abrikosov lattice), where the lattice constant is of the order of the oscillator length $a_0$. 

With the help of our code we perform a set of numerical experiments and obtain the trend of variation of vortex number with other free parameters. 

\begin{figure}[]
\centering 
\includegraphics[width=0.9\textwidth, height=0.7\textwidth]{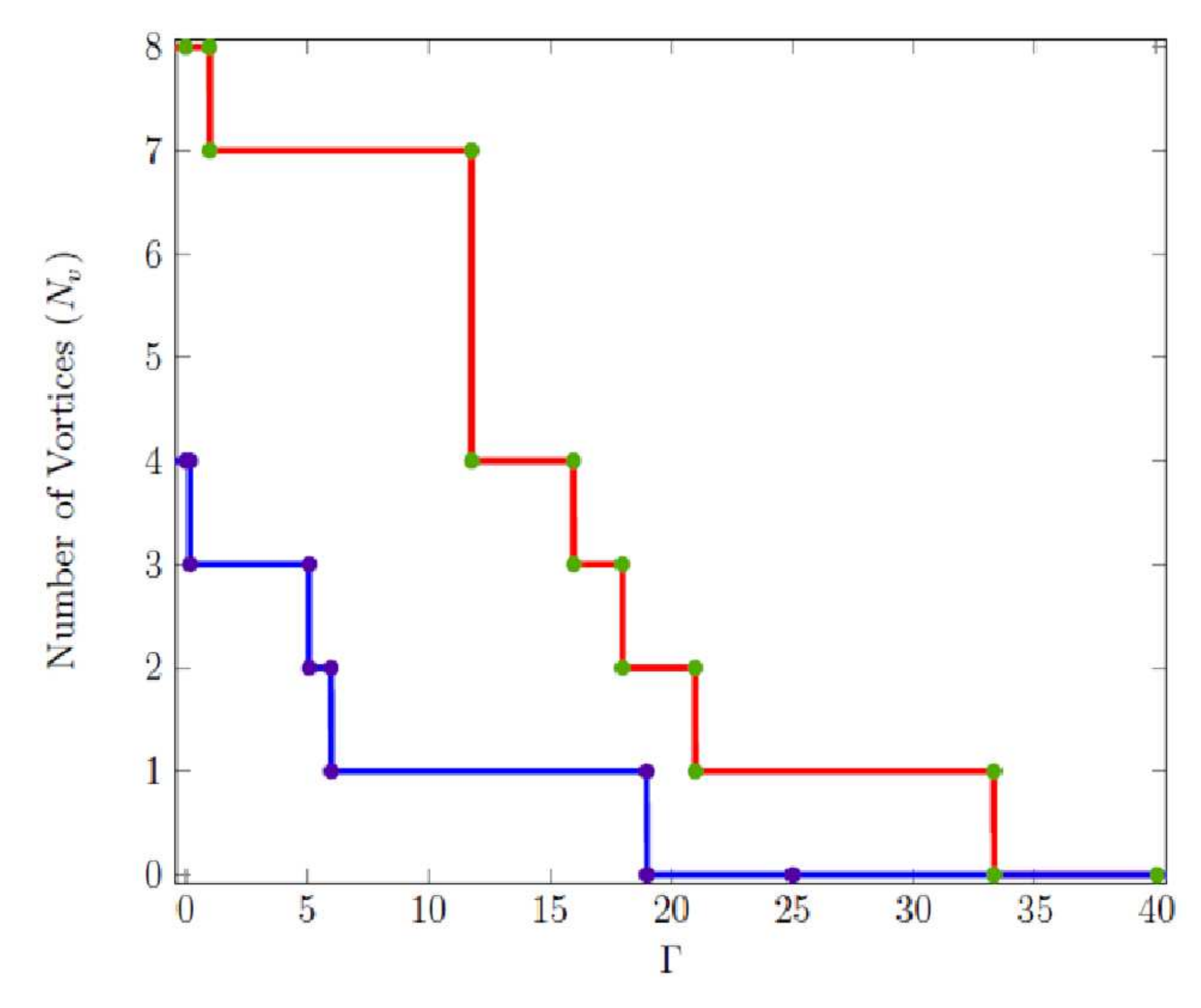} 
\caption{Variation of number of vortices $N_v$ with $\Gamma$ are shown by the solid red ($\Omega$ = 0.9) and blue ($\Omega$ = 0.8) lines. The variation is a step function since the vortex {\it number} can not be non-integer. Condensate size is $16a_0$, $g=50$, $\gamma_{eff}=20$.}
\label{NvGam}
\end{figure}

{\it Results for varying $\Gamma$}: We find that the number of vortices (for fixed $g, \Omega$) {\it decreases} as the value of $\Gamma$ increases (see Fig. \ref{NvGam}) and ultimately, after a certain value, the vortices. And ultimately after a certain value the vortices completely disappear. This value of $\Gamma$ is referred to as the {\it critical gamma} or $\Gamma^0$. It might happen that there are few vortices on the edge but as the condensate rotates these vortices continue forming and disappearing with time. This will be a discussed in the next section (see Fig. \ref{NoVort}). Since this critical value of gamma exists, we emphasize that one can realize a vortex free condensate and hence realize the synthetic magnetic field on such systems.

\begin{figure}[]
\centering
\subfloat[]{
\includegraphics[width=0.5\textwidth, height=0.4\textwidth]{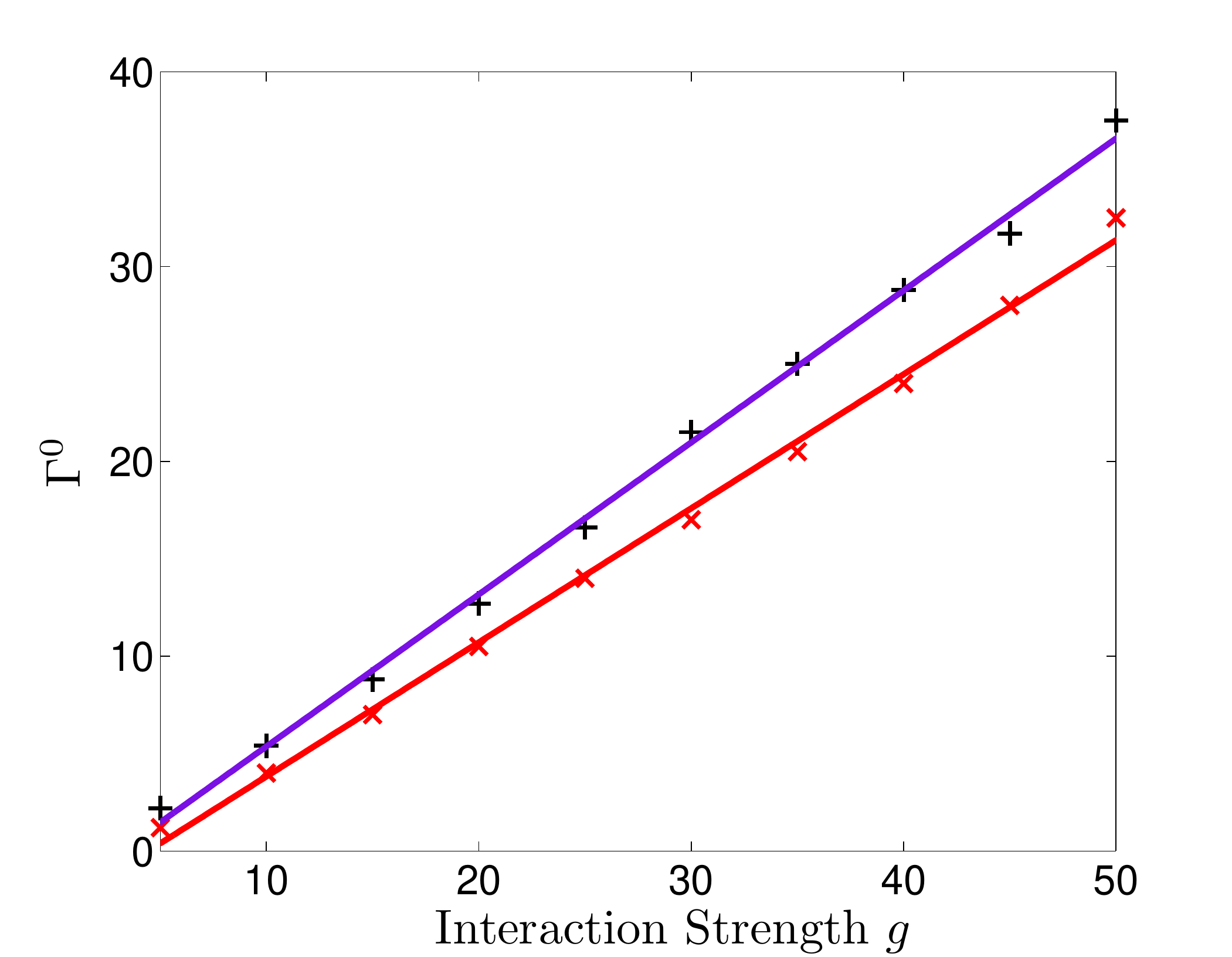} \label{gvsCrit} }
\subfloat[]{
\includegraphics[width=0.5\textwidth, height=0.4\textwidth]{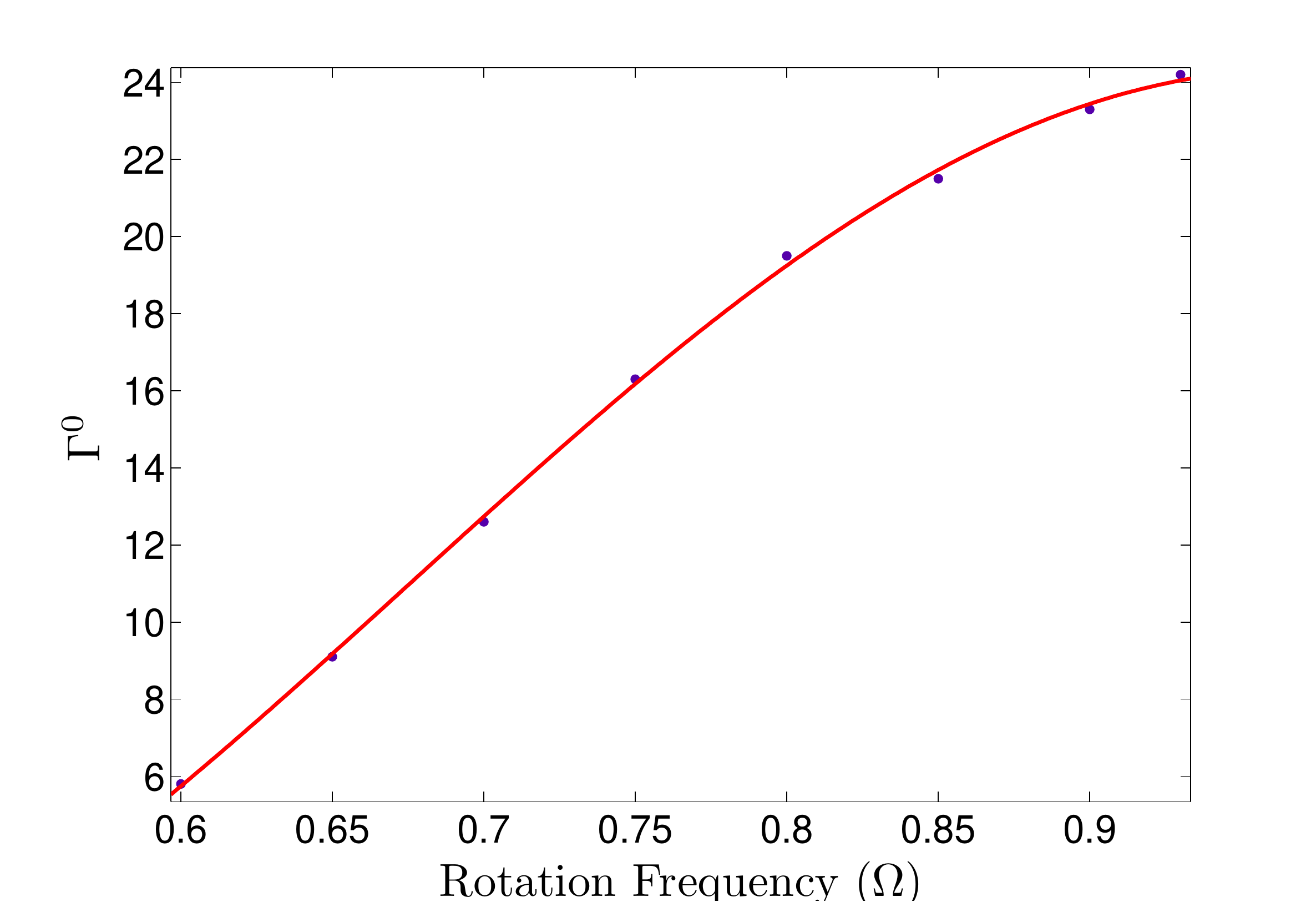} \label{OmvsCrit} }
\caption{Variation of $\Gamma^0$ with: (a) inter-polariton interaction strength ($g$). The two lines correspond to $\Omega$ = 0.90 (red) and 0.96 (blue); (b) variation with rotation frequency ($g = 40$, $\gamma_{eff}=20$).}
\label{Crit}
\end{figure}

This behavior can be understood from the hydrodynamic equations of the MCP superfluid. We will provide its full justification immediately after we derive the steady-state hydrodynamic equations for the cGPE in the next section. Now we present another observation: how the value of $\Gamma_0$ varies with $g$ or $\Omega$. These variations are shown in Fig. \ref{Crit}.

\begin{figure}[]
\centering
\subfloat[]{
\includegraphics[width=0.5\textwidth, height=0.4\textwidth]{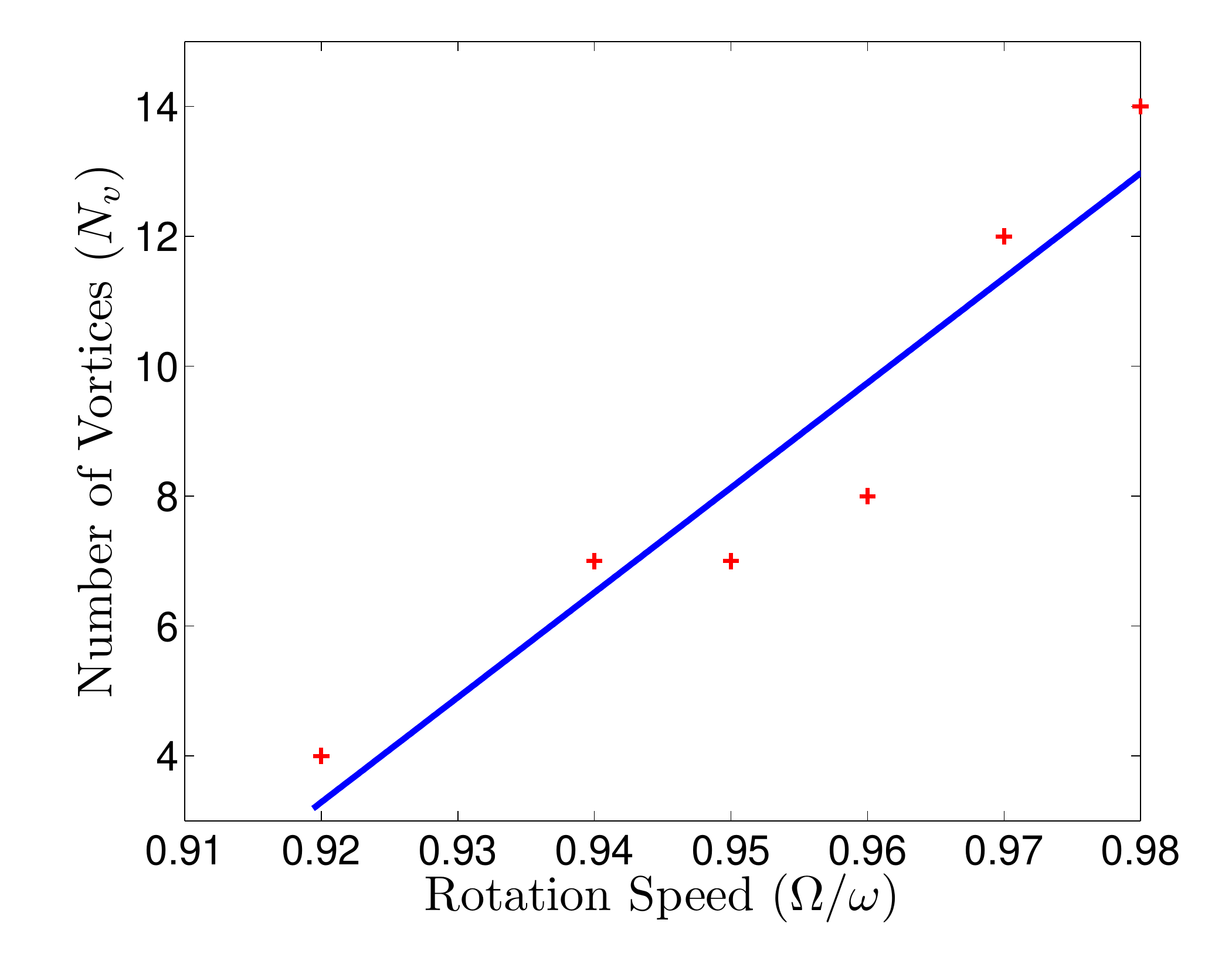} \label{OmvsNmax} }
\subfloat[]{
\includegraphics[width=0.5\textwidth, height=0.4\textwidth]{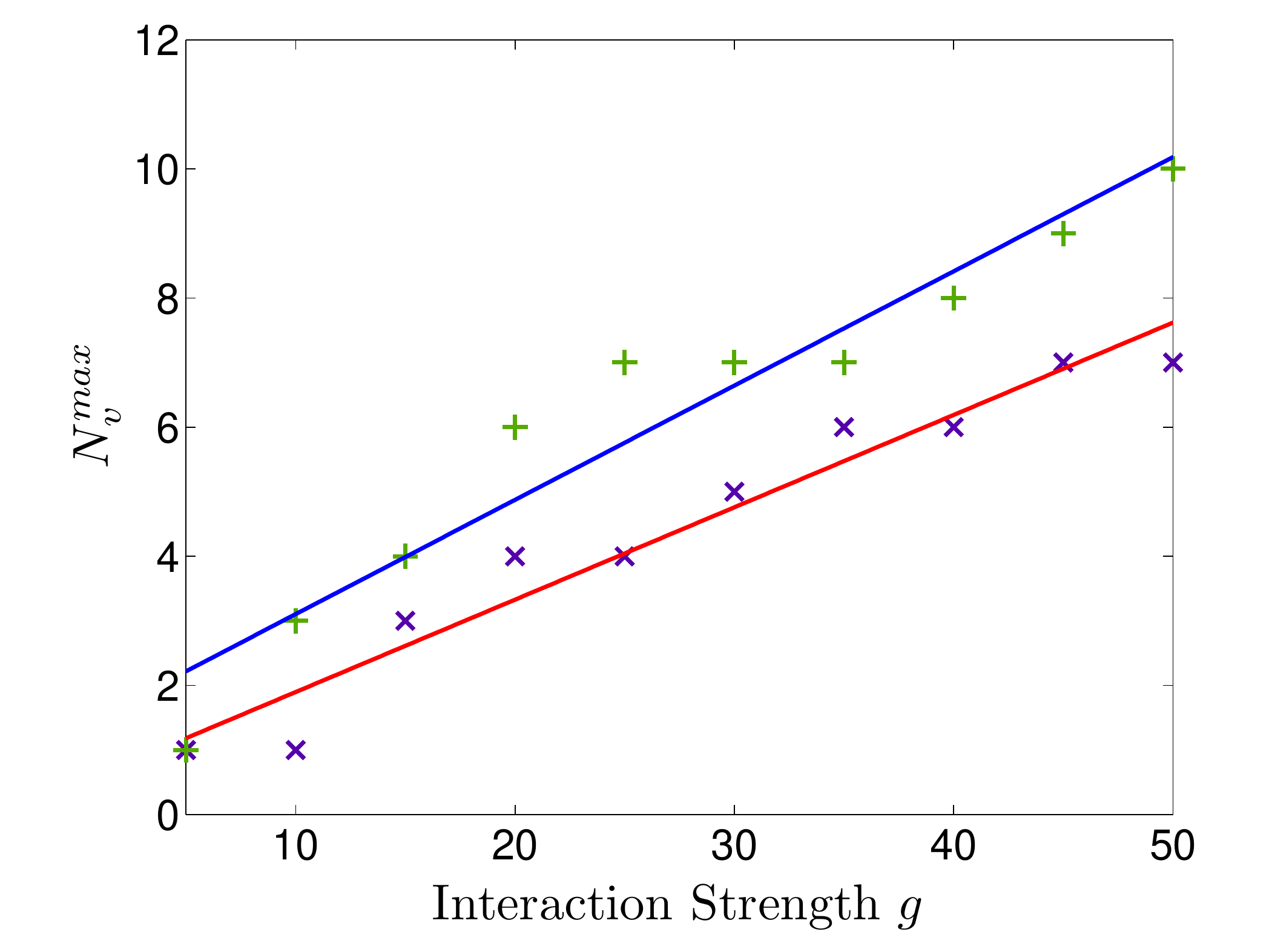} \label{gvsNmax} }
\caption{(a) Variation of number of vortices with rotation frequency ($\Gamma=16, \gamma_{eff}=20, g=50$). Blue line corresponds to the average variation; (b) variation of maximum number of vortices formed ($N_v^{max}$, that is when $\Gamma$ is set to zero, refer to Figure \ref{NvGam}), with inter-polariton interaction strength, $g$. The red and blue lines are average variations for rotation frequencies $\Omega$ = 0.93, 0.96, respectively.}
\label{evolve0}
\end{figure}

{\it Result for varying $\Omega$}: It is a well established fact \cite{Stringari} that the number of vortices increases as the rotation frequency increases, or the interaction strength increases. In case of MCP, this also holds true. Hence, what we verified is, how the $N_v^{\max}$ depends on $\Omega$, see Fig. \ref{OmvsNmax}. This is trivially a consequence of the fact that $N_v$ increases with $\Omega$. It may be noted that, since we are mostly interested in critical rotation, we restrict our study to $\Omega \simeq 1$. 

{\it Result for varying $g$}: Similar to the previous case, we study the variation of $N_v^{max}$ with $g$, see Fig. \ref{gvsNmax}. As expected $N_v^{max}$ increases with $g$ \cite{Fetter}.

The mean vortex density increases with the rotation frequency or inter-particle interaction \cite{Feynman}. From mean-field calculations it can also be shown that the separation between the vortices decreases as the rotation frequency increases \cite{Fetter2}. Hence one can expect the total number of the vortices in the condensate to increase with increasing rotation speed or interaction strength. 
 
\section{Time-evolution}
\label{sec:ResultEvo}

We solve the the full time-dependent cGPE in Eq. \eqref{cGPE} using GPELab (the method briefed in appendix \ref{app:gpelab} and further details on dynamic calculations can be found in \cite{Antoine2014}). In order to understand the time-evolution of the MPC to it's equilibrium state we first derive the hydrodynamic equations. Considering the system to be large enough such that the Thomas-Fermi approximation applies, the equation of motion of the condensate can be well described by the hydrodynamics theory of superfluids \cite{hydro}. A similar equation of motion of the polariton superfluid, in a frame rotating with angular velocity $\Omega \hat{z},$ can be derived from Eq. \eqref{cGPE},
\bea
\frac{\partial n_c}{\partial t} + \nabla \cdot [ n_c (\bs v_s - \bs \Omega \times \bs r)] = 2 (\gamma_{eff} - \Gamma n_c) n_c/ \hbar, \label{BernoulliI} \\
\frac{\partial \bs v_s}{\partial t} + \nabla \Big [ \frac{v_s^2}{2} + \frac{V}{m} + \frac{gn_c}{m} - \bs v_s . (\bs \Omega \times \bs r) \Big ] = 0.
\label{BernoulliII}
\eea
Eq. \eqref{BernoulliI} is the continuity equation and (\ref{BernoulliII}) is Bernoulli's equation. The derivations are provided in the Appendix \ref{app:hydro}. Assuming the gas to be dilute we have replaced the density of the gas, $|\Psi(\bs r, t)|^2$ with the condensate density, $n_c(\bs r, t)$. The equilibrium state solutions in the rotating frame can be obtained by setting $ \partial n/ \partial t = \partial v/ \partial t = 0.$ Moreover, the behavior of the condensate can be explained from these hydrodynamics equations.

\begin{figure}[]
\centering
\includegraphics[width=0.8\textwidth, height=0.55\textwidth]{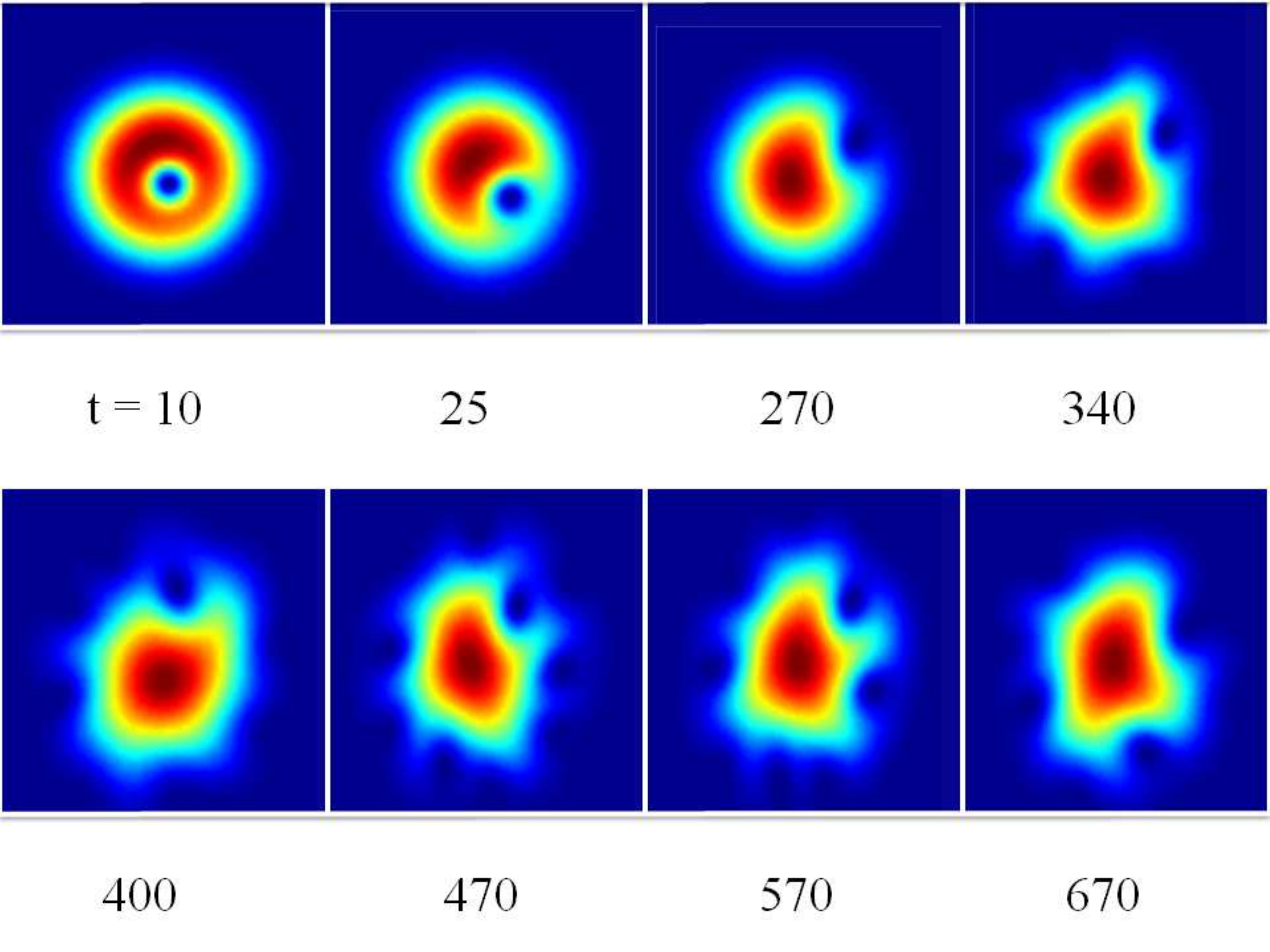} \caption{The process of condensate evolution till an equilibrium state is achieved. The time taken (in the unit of $1/\omega_\perp$) is written corresponding to every snapshot. With time the vortex gradually comes towards the edge of the condensate and ultimately the central part of the condensate becomes vortex free. The value of other parameters are $\Omega$ = 0.95, $g$ =30, $\Gamma$ = 20 ($ > \Gamma^0$), $\gamma_{eff}$ = 20. The size of the condensate is $16a_0$. }
\label{NoVort}
\end{figure}

Let us now explain the existence of $\Gamma^0$ using these equations: in the equilibrium state the linear imaginary part and the non-linear imaginary part of the cGPE become equal, see \eqref{BernoulliI}. Hence if $\Gamma$ increases, keeping all other parameters fixed, then $|\Psi|^2$ decreases or the number of particles decreases, i.e. the condensate shrinks. Hence the vortices start vanishing. We illustrate this process in Fig. \ref{NoVort}, where we rotate the condensate with a critical gamma. 

Furthermore, when $\Gamma$ is set to $\Gamma_0$, the condensate reaches the critical size needed to expel all the vortices to its boundary. As we know, the number of vortices increases with increasing $\omega$, or $g$, hence, one has to increase the value of $\Gamma^0$ to obtain a vortex-free condensate. This result is in agreement with what we observe in Fig. \ref{Crit}. 

\begin{figure}[]
\centering
\includegraphics[width=0.85\textwidth, height=0.6\textwidth]{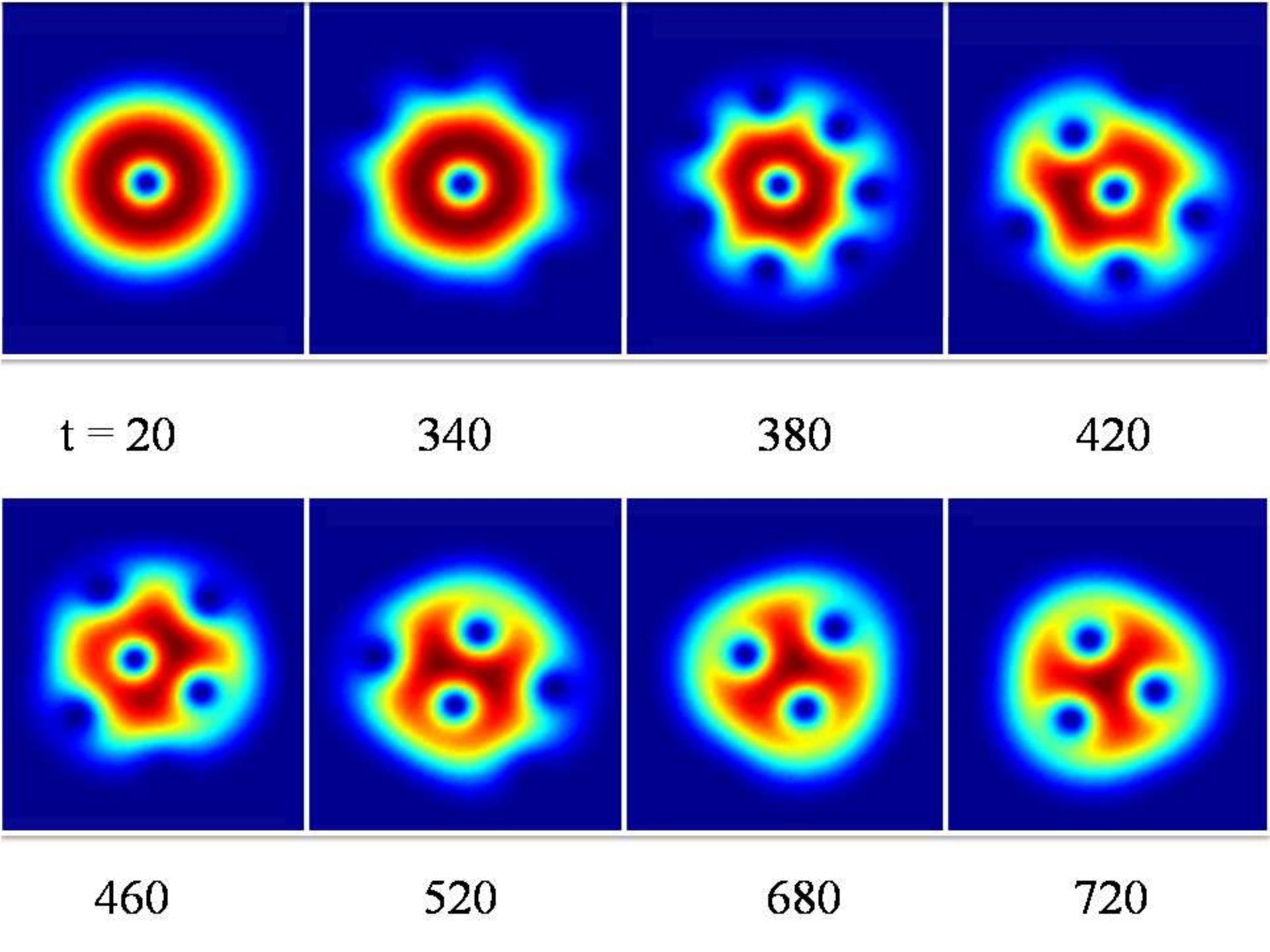} 
\caption{Evolution of the condensate until an equilibrium state is reached. The time taken (in the unit of $1/\omega_\perp$) is written corresponding to every snap. With time the vortex gradually comes towards the edge of the condensate and ultimately the central part of the condensate becomes vortex free. The value of other parameters are $\Omega$ = 0.92, $g$ =40, $\Gamma$ = 15, $\gamma_{eff}$ = 20. The size of the condensate is $16a_0$.}
\label{evolve3}
\end{figure}

{\it Results for $\gamma_{eff}$}: As mentioned before, $\gamma_{eff}$ does not affect the steady-state but it affects the time-evolution to the steady-state. Considering the damping term $\gamma_{eff}$ not equal to zero, we can easily deduce from Eq. \eqref{BernoulliI} that the condensate will take more time to evolve to a steady-state as the value of $\gamma_{eff}$ decreases. In Fig. \ref{evolve3}, we illustrate the evolution of a condensate for a finite value of the $\gamma_{eff}$. 

\section{Conclusion}
\label{sec:Conclusion}

Now we summarize our main results. We found that in the equilibrium state, the MPC always forms triangular vortex lattice (see Fig. \ref{Vortices}). For the MPC, we re-established that the number of vortices, $N_v$ increases as the rotation frequency, $\Omega$ or the interaction strength increases (Fig. \ref{Crit}). Our primary result is that, we show, for a given rotation frequency there exists a critical value of $\Gamma$, beyond which the condensate becomes vortex free (see Fig. \ref{NoVort}). We explained this result using the hydrodynamic theory of superfluid. Further, this critical value can also be controlled by tuning rotation frequency or the interaction strength. We hope this result will be helpful for experimentalists to tune the right range of parameters for the MPC such that synthetic magnetic field can be realized.

There are certain questions which are yet to be addressed. Such as to see how (the time scale in which) the condensate will relax if we suddenly switch of the (a) trap potential, or (b) the rotation. As a natural extension, one can consider asymmetric trap potentials and study how the anisotropy will provide a directional preference to the condensate dynamics. Many of these studies can be accomplished through the GPELab, with certain amount of modifications. 

\section{Acknowledgment}
B. P. thanks J. Keeling for helpful comments on the manuscript and I. Carusotto, S. Ghosh for helpful discussions. R. D. acknowledges this work was partially supported by the French ANR grant MicroWave NT09 460489 (“Programme Blanc” call) and ANR-12-MONU-0007-02 BECASIM (Mod\`eles Num\'eriques call).

The work was planned by B.P. and R.S. The GPELab package was developed by R.D. The parent code was extended for polaritons and various analysis were performed by B.P. and A.N. 


\clearpage
\newpage
\pagebreak
{\centerline {\Large {\bf APPENDIX}} }
\appendix

\section{GPELab} 
\label{app:gpelab}

In order to launch a computation with GPELab, we have to write a script to define the parameters of the scheme and the definition of the physical problem. We now show how we wrote the scripts that are used in this paper. Before launching the simulation, we have to define three structures: the \texttt{Method} ( resp. the \texttt{Geometry2D}) structure that includes all the parameters concerning the method ( resp. the geometry) of the computation and the \texttt{Physics2D} structure where the operators of the \textit{complex} Gross-Pitaevskii equation have to be defined. In our computations, we choose a time step equal to $0.1$ and a stopping criterion $\varepsilon = 10^{-12}$ in the parameters of the \textit{imaginary time} method. This is coded in Table \ref{MethodVar} where we define the \texttt{Method} structure by using the \texttt{Method\_Var2d} function.

\begin{table}[h!]
\fbox{\begin{minipage}{1\textwidth}
\texttt{Computation = 'Ground';}
\\ \texttt{Ncomponents = 1;}
\\ \texttt{Type = 'BESP'; }
\\ \texttt{Deltat = 1e-1;}
\\ \texttt{Stop\_time = [];}
\\ \texttt{Stop\_crit = \{'Energy', 1e-12\};}
\\ \texttt{Method = Method\_Var2d(Computation,Ncomponents, Type, Deltat, Stop\_time, Stop\_crit);}
\end{minipage}}
\caption{Defining the \texttt{Method} structure. }\label{MethodVar}
\end{table}

Next, we define the \texttt{Geometry2D} structure. The computational box is set to $\mathcal{O} = [-15,15]^2$. We fix a number of grid points of $2^9+1$ in the $x$- and $y$-directions. 

We now have to set the physical problem. That is, we have to define each operators of cGPE: the Laplace operator $-\frac 1 2 \nabla^2$, the potential operator $V(\boldsymbol{x})$, the rotational operator $\Omega L_z$, and the nonlinear operator $g|\Psi|^2 + i\left[\gamma_{\textrm{eff}} - \Gamma|\Psi|^2\right]$.

GPELab is able to handle general Gross-Pitaevskii equations and includes functions that define several types of operators \cite{Antoine2014}. That is, we can compute the stationary state of a nonlinear Schr\"odinger equation of the form
\begin{equation}
i\partial_t \Psi = D(-i\nabla)\Psi + \sum_{j = 1}^d G_j(\boldsymbol{x})\partial_{\boldsymbol{x}_j} \Psi + V(\boldsymbol{x})\Psi + F(\Psi,\boldsymbol{x})\Psi,
\end{equation}
where $D$ is a dispersion operator, $G_j \partial_{\boldsymbol{x}_j}$ are gradient operators, $V$ is a potential operator and $F$ is a nonlinear operator. In Table \ref{PhysicsVar}, we set the operators of cGPE by using the \texttt{Dispersion\_Var2d}, \texttt{Potential\_Var2d}, \texttt{Nonlinearity\_Var2d}, \texttt{Gradientx\_Var2d} and \texttt{Gradienty\_Var2d} functions (see \cite{Antoine2014} for more details). Furthermore, we fix here, for example, the parameters $g = 50$, $\gamma_{\textrm{eff}} = 5$, $\Gamma = 50$, $\Omega = 0.5$ and $\gamma_x = \gamma_y = 1$.

\begin{table}[h!]
\fbox{\begin{minipage}{1\textwidth}
\texttt{Physics2D = Physics2D\_Var2d(Method,1,1);}
\\\texttt{Physics2D = Dispersion\_Var2d(Method, Physics2D, ...}
\\\texttt{ @(fftx,ffty) (1/2)*(fftx.\textasciicircum2+ffty.\textasciicircum2));}
\\\texttt{Physics2D = Potential\_Var2d(Method, Physics2D, ...}
\\\texttt{@(x,y) (1/2)*(gamma\_x.*x.\textasciicircum2+gamma\_y.*y.\textasciicircum2));}
\\\texttt{Physics2D = Nonlinearity\_Var2d(Method, Physics2D, ...}
\\\texttt{@(phi,x,y) g*abs(phi).\textasciicircum2 + 1i*(gamma\_eff - Gamma*abs(phi).\textasciicircum2) );}
\\\texttt{Physics2D = Gradientx\_Var2d(Method, Physics2D,@(x,y) 1i*Omega*y);}
\\\texttt{Physics2D = Gradienty\_Var2d(Method, Physics2D,@(x,y) -1i*Omega*x);}
\end{minipage}}
\caption{Defining the \texttt{Physics2D} structure. }\label{PhysicsVar}
\end{table}

The initial data is computed with the help of the \texttt{InitialData\_Var2d} provided in GPELab. Before launching the computation, we have to define additional structures that we do not detail here. These structures are used to define the outputs computed during the simulation (\textit{e.g.} drawing the solution, computing quantities such as the energy, the angular momentum, etc.). At the end of the simulation, we can proceed to extract the outputs and analyze the data. 


\section{Hydrodynamic theory of cGPE} 
\label{app:hydro}

In this appendix we show that the amplitude part and the phase part of the cGPE in Eq. \eqref{cGPE} can be decoupled into two hydrodynamic equations, {\it viz.} the equation of continuity (EoC) and equation of motion (EoM). For simplicity we write $\Psi(\bs r, t)$ as $\Psi$. The cGPE of our consideration is 
\beq
\imath \hbar \frac{\partial \Psi}{\partial t} = \Big [ -\frac{\hbar^2 \nabla ^2}{2m} + V(\bs r) + g|\Psi|^2 + \imath \hbar \nabla . (\bs \Omega \times \bs r) +  \imath (\gamma_{eff} - \Gamma |\Psi|^2) \Big] \Psi. 
\label{SupcGPE}
\eeq
Notice we have used the vector identity $ \nabla . (\bs \Omega \times \bs r) = \bs r . (\nabla \times \bs \Omega) + \bs \Omega . (\bs r \times \nabla).$ Since the frame is rotating with a {\it constant} angular velocity, $\bs \Omega = \Omega \hat{z},$ the first term on the right hand side vanishes. Now we show the real part of this equation is the EoC and the imaginary part is equivalent to the EoM or Bernoulli's equation.

{\it Equation of Continuity} : Multiplying $\bar{\Psi}$ on both sides of the equation \eqref{SupcGPE} and subtracting its conjugate we get 
\beq
\imath \hbar \frac{\partial \Psi \bar{\Psi}}{\partial t} = -  \frac{\hbar^2}{2m} \Big ( \bar{\Psi} \nabla ^2 \Psi - \bar{\Psi} \nabla ^2 \Psi \Big )+ \imath \hbar \nabla . (\bs \Omega \times \bs r) |\Psi|^2 + 2\imath (\gamma_{eff} - \Gamma |\Psi|^2) |\Psi|^2. 
\eeq
The first term on the right hand side can be identified with divergence of fluid current density, $\nabla \cdot \bs j(\bs r, t),$ where $\bs j = -i \hbar \big ( \bar{\Psi} \nabla  \Psi - \bar{\Psi} \nabla \Psi \big )/2m$. Assuming the gas to be dilute we can identify density of the gas, $|\Psi(\bs r, t)|^2$ with the condensate density, $n_c(\bs r, t)$ 
\bea
\frac{\partial n_c}{\partial t} + \nabla \cdot \bs j - \nabla . (\bs \Omega \times \bs r) n_c= 2 (\gamma_{eff} - \Gamma n_c) n_c / \hbar
\eea
Using Madelung transformation, $\Psi (\bs r, t) = \sqrt{n_c(\bs r, t)} e^{\imath \phi(\bs r, t)}$, the order parameter can be expressed in terms of its amplitude and phase factor: $j (\bs r, t) = n_c (\bs r, t) \bs v_s (\bs r, t),$ $\bs v_s (\bs r, t) = \hbar \nabla \phi (\bs r, t) / m$. Here, $\bs v_s (\bs r, t)$ is the velocity of fluid flow, which is irrotational ( $\nabla \times \bs v_s = 0$ ). The subscript in the velocity stands for 'superfluid', as irrotational flow is one of the characteristics of superfluids. So the equation of continuity for a rotating condensate becomes
\beq
\frac{\partial n_c}{\partial t} + \nabla .[ n_c (\bs v_s - \bs \Omega \times \bs r)] = 2 (\gamma_{eff} - \Gamma n_c) n_c/ \hbar. 
\eeq

{\it Bernoulli's Equation of Motion}: We express the order parameter in terms of amplitude and phase and rewrite the left hand side of the cGPE as
\beq
\imath \hbar \partial_t (\sqrt{n_c}e^{\imath \phi}) = \frac{\imath \hbar e^{\imath \phi}}{2\sqrt{n_c}} \partial_t n_c - \hbar \sqrt{n_c} e^{\imath \phi} \partial_t \phi .
\eeq
The EoC and the cGPE can be used to simplifying this further to
\bea
\frac{\imath \hbar e^{-\imath \phi}}{\sqrt{n_c}} \nabla . (\bs \Omega \times \bs r) \sqrt{n_c}e^{\imath \phi} + \frac{\imath \hbar }{2n_c} \nabla .[ n_c (\bs v_s - \bs \Omega \times \bs r)] = \nn \\ \frac{\imath \hbar^2 }{2mn_c} \nabla n_c \nabla \phi + \frac{\imath \hbar^2 }{2m} \nabla^2 \phi - \hbar \nabla \phi . (\bs \Omega \times \bs r) .
\eea
Inserting the above two equations we finally obtain (rewriting this equation by using $ \nabla \phi = m \bs v_s /\hbar, $ and taking divergence of both sides of the equation)
\beq
m \frac{\partial \bs v_s}{\partial t} + \nabla \Big [ \frac{m v_s^2}{2} + V + gn_c - m \bs v_s . (\bs \Omega \times \bs r) - \frac{\hbar^2 }{2m} \frac{\nabla^2 \sqrt{n_c}}{\sqrt{n_c}} \Big ] = 0.
\eeq
The last term, called the 'quantum pressure' \cite{Stringari} term, has important significance for inhomogeneous gases. However, if the density of the gas changes slowly in space, then this term can be neglected. In the ground state of the condensate the pressure term scales almost as the inverse square of the size of the cloud. In our case we assume the size of the condensate is much larger than the healing length $\xi = \hbar / \sqrt{2mgn_c}$, hence drop the pressure term (Thomas-Fermi approximation) and obtain
\beq
\frac{\partial \bs v_s}{\partial t} + \nabla \Big [ \frac{v_s^2}{2} + \frac{V}{m} + \frac{gn_c}{m} - \bs v_s . (\bs \Omega \times \bs r) \Big ] = 0.
\eeq
This is the Bernoulli's equation of hydrodynamics. It must be noted that, Thomas-Fermi approximation is equivalent to ignoring the kinetic term in the original cGPE.


\end{document}